\documentclass[twocolumn,showpacs,pra,aps,superscriptaddress]{revtex4}
\usepackage{ulem}
\usepackage{array}
\usepackage{amsmath}
\usepackage{graphicx}
\begin{document}

\title{Greenberger-Horne-Zeilinger Nonlocality in Arbitrary Even
  Dimensions}

\author{Jinhyoung Lee} 

\affiliation{Department of Physics, Hanyang University, Seoul 133-791,
  Korea}

\affiliation{Quantum Photonic Science Research Center, Hanyang
  University, Seoul 133-791, Korea}

\author{Seung-Woo Lee} 

\affiliation{Quantum Photonic Science Research Center, Hanyang
  University, Seoul 133-791, Korea}

\author{M. S. Kim}

\affiliation{School of Mathematics and Physics, The Queen's
  University, Belfast, BT7 1NN, United Kingdom}

\date{\today}

\begin{abstract}
  We generalize Greenberger-Horne-Zeilinger (GHZ) nonlocality to every
  even-dimensional and odd-partite system. For the purpose we employ
  concurrent observables that are incompatible and nevertheless have a
  common eigenstate. It is remarkable that a tripartite system can
  exhibit the genuinely high-dimensional GHZ nonlocality.
\end{abstract}


\maketitle

\section{introduction}
\label{sec:int}

Quantum nonlocality is one of the most profound virtues inherent in
quantum mechanics and it is a fundamental resource for quantum
information processing. Quantum nonlocality is implied by the Bell
theorem that quantum mechanics conflicts with any local realistic
theories. In a bipartite system, Bell constructed a statistical
inequality that all local realistic theories should satisfy but quantum
mechanics can violate \cite{Bell64}. Later, Greenberger, Horne, and
Zeilinger (GHZ) proved the Bell theorem without any inequalities in a
tripartite system \cite{GHZ89}. The so-called GHZ state that exhibits
the GHZ nonlocality has been employed as a quantum channel for quantum
key distribution \cite{Kempe99} and quantum secret sharing
\cite{Hillery99}. For complex tasks of scalable quantum computation and
quantum error correction, nature of multipartite entanglement and its
nonlocality test have become important issues.

During the early period, discussions on nonlocality were centered at
two-dimensional systems such as spins and polarizations. However, while
most physical systems are defined in higher-dimensional Hilbert spaces,
only little is known about the higher-dimensional multipartite systems.
Bell's inequality has been generalized to an arbitrary-dimensional
bipartite system \cite{Kaszlikowski00}. Very recently, Bell's inequality
was discussed for a three-dimensional tripartite system \cite{Acin04}.
In order to generalize GHZ nonlocality to an arbitrary even dimensional
system, \.Zukowski and Kaszlikowski suggested an experiment using
optical elements such as multiport beam splitters and phase shifters
\cite{Zukowski99}. Their work was compensated by Cerf {\em et al.} with
Mermin's formulation which emphasizes relations between a set of
operators \cite{Cerf02,Mermin90}. The works by the both groups require
$N$ subsystems, with $N=d+1$, to exhibit the $d$-dimensional GHZ
nonlocality for $d$ an even integer. GHZ-like nonlocality with
statistical expectations was argued for a $d$-dimensional $d$-partite
system \cite{Kaszlikowski02}. On the other hand, the original GHZ
nonlocality requires only three subsystems in two-dimensional Hilbert
space.  Here an extremely important question arises.  Is there no such a
nonlocality test without inequalities for a $d$-dimensional $N$-partite
system where $N$ is independent of $d$? We answer this question in this
paper.

\.Zukowski and Kaszlikowski \cite{Zukowski99} and Cerf {\em et al.}
\cite{Cerf02} started their arguments from the compatible composite
observables which led the discussion to the local complementary
observables. This approach is derived from GHZ's original study.
However, we go back to the argument on physical reality by Einstein,
Podolsky, and Rosen (EPR) \cite{Einstein35}. We will show that this
leads to a generalization of GHZ nonlocality which is used for a nonlocality
test in a multi-dimensional multipartite system {\em without statistical
  inequalities}.

In this paper, to prove the Bell theorem without inequalities, we employ
concurrent observables, which are mutually incompatible but still have
their common eigenstate. The concurrent observables are chosen such that
the common eigenstate is ``a generalized GHZ state.''  Their local
observables are shown to be involved in elements of physical reality
according to EPR's criterion \cite{Einstein35}. It is proved that our
generalization is genuinely multi-dimensional. We emphasize that this
work first shows that a {\em tripartite} system suffices for the
genuinely $d$-dimensional GHZ nonlocality with $d$ an even integer. We
discuss its extension to a multipartite system.

\section{concurrent observables and elements of physical reality}
\label{sec:epr}

EPR \cite{Einstein35} proposed as a sufficient condition for recognizing
an element of physical reality, ``If, without in any way disturbing a
system, we can predict with certainty the value of a physical quantity,
then there exists an element of physical reality corresponding to this
physical quantity.'' Elements of physical reality accompanied with
Einstein's locality play an essential role in nonlocality.  They have
been investigated by finding compatible observables and their common
eigenstate for a given composite system.  For instance, in the EPR-Bohm
paradox of two spin-1/2 particles \cite{Bohm51}, Bohm considered the set
of commuting observables,
$\{\hat{\sigma}_a\otimes\hat{\sigma}_a | a=x,y,z\}$, and their
simultaneous eigenstate, $|\psi\rangle = (|\uparrow \downarrow\rangle -
|\downarrow \uparrow\rangle)/\sqrt{2}$.  Finding the compatible
observables has been regarded as a crucial step to recognize the
elements of physical reality.  The approach of the compatible
observables (for elements of physical reality) can be extended when one
faithfully follows EPR's criterion.  The extension is one of the key
points in generalizing GHZ nonlocality.

For a quantum system of $d(>2)$ dimension, there are some incompatible
observables which nevertheless have a common eigenstate \cite{Peres98}.
The observables whose common eigenstate is equal to a given system state
are called {\em concurrent observables}. The measurement results for the
concurrent observables can simultaneously be specified as far as the
quantum system is prepared in their common eigenstate. Note that
compatible observables are concurrent observables if the quantum system
is prepared in one of their common eigenstates. Following EPR's
criterion, one can involve concurrent observables (more precisely, their
local observables) in elements of physical reality. Here it is crucial
that the system state is an eigenstate of the composite observables. For
instance, a composite system of subsystems $A$ and $B$ is prepared in a
quantum state $|\psi\rangle$ and the two subsystems are separated at a
long distance.  The state $|\psi\rangle$ is chosen such that it is an
eigenstate of a composite observable $\hat{X} \otimes \hat{Y}$: $\hat{X}
\otimes \hat{Y} |\psi \rangle = \lambda |\psi \rangle$ where $\lambda$
is the corresponding eigenvalue.  Suppose that the variable $X$ for the
subsystem $A$ is measured and its outcome is $x$ (one of the eigenvalues
of $\hat{X}$). One can predict with certainty the value of $Y$, i.e. $y
= \lambda/x$, for the subsystem $B$.  Assuming Einstein's locality, as
the two systems are separated in a long distance, the measurement
performed on the subsystem $A$ can instantaneously cause no real change
in the subsystem $B$. Thus, the variable $Y$ is predetermined {\em
  before the measurement} and it is an element of physical reality
according to EPR's criterion.  Similarly, the variable $X$ is also an
element of physical reality.

It is in general difficult to find all concurrent observables.  Instead,
a particular set of them is easily found once symmetries are known for a
given quantum state. Suppose that a quantum state $|\psi\rangle$ of a
given system is an eigenstate of an observable $\hat{X}$ with the
eigenvalue $\lambda$: $\hat{X} |\psi\rangle = \lambda |\psi\rangle$. Let
$G$ be a group of symmetry operations such that each operation $g \in G$
is represented by some unitary operator $\hat{V}(g)$ which leaves the
quantum state invariant, i.e., $\hat{V}(g)|\psi\rangle = |\psi\rangle$.
Then the quantum state $|\psi\rangle$ is the common eigenstate, with the
same eigenvalue $\lambda$, of the composite observables
$\hat{X}(g)=\hat{V}(g)\hat{X}\hat{V}^\dag(g)$:
\begin{eqnarray}
  \label{eq:co}
  \hat{X}(g) |\psi\rangle = \hat{V}(g)\hat{X}\hat{V}^\dag(g)|\psi\rangle
  = \lambda |\psi\rangle.
\end{eqnarray}
Here the form of the unitary operator $\hat{V}(g)$ was not conditioned.
However, in order to discuss on elements of physical reality, we require
that such a unitary operator should be in the form of the tensor product
of local unitary operators: For instance,
$\hat{V}(g)=\hat{U}_1(g)\otimes \hat{U}_2(g)$ for a bipartite system.

Consider a tripartite system of $A$, $B$, and $C$. Each subsystem is of
$d$ dimension, hence called qudit. The composite system is assumed to be
in a state,
\begin{eqnarray}
  \label{eq:ghzs}
  |\psi\rangle = \frac{1}{\sqrt{d}} \sum_{n=0}^{d-1} |n,n,n\rangle,
\end{eqnarray}
where $\{|n\rangle\}$ is a complete orthonormal basis set. The state
$|\psi\rangle$ is conventionally called a generalized GHZ state.  Let us
consider a unitary operator in the form of
\begin{eqnarray}
  \label{eq:lut}
  \hat{V} = \hat{U}_1 \otimes \hat{U}_2 \otimes \hat{U}_3,
\end{eqnarray}
where
\begin{eqnarray}
  \label{eq:lust}
  \hat{U}_\alpha = \sum_{n=0}^{d-1} \omega^{f_\alpha(n)}
  |n\rangle\langle g(n)|.
\end{eqnarray}
Here $\omega = \exp(2\pi i /d)$ is a primitive $d$-th root of unity over
complex field and $g(n)$ is a permutation map on the set
$D=\{0,1,\cdots,d-1\}$. Note that $\hat{U}_\alpha$ reduces to a phase
shift operator if $g(n)$ is an identity map. The unitary operator
$\hat{V}$ leaves $|\psi\rangle$ invariant if
\begin{eqnarray}
  \label{eq:pfluoiqs}
  f_1(n) + f_2(n) + f_3(n) \equiv 0 \mod d,
\end{eqnarray}
for each $n \in D$. The expression of ``$x \equiv y \mod d$'' implies
that $(x-y)$ is congruent to zero modulo $d$ throughout the paper.

\section{generalized GHZ nonlocality}
\label{sec:ngghz}

\subsection{Tripartite system}
\label{sec:ts}

Suppose that three observers, say, Alice, Bob, and Charlie are mutually
separated at a long distance and they will perform their measurements on
the qudits $A$, $B$, and $C$, respectively. Each observer is allowed to
choose one of two variables, $X$ and $Y$.  The choice is made by
deciding local parameters in each measuring device. Each variable takes,
as its value, an element in the set of order $d$, $S=\{1, \omega, \dots,
\omega^{d-1}\}$. The elements of $S$ are the $d$-th roots of unity over
the complex field.

In quantum mechanics, an {\em orthogonal} measurement is described by a
complete set of orthonormal basis vectors, $\{|n\rangle_{p}\}$, where
the subscript $p$ denotes the set of parameters in the measuring device.
Distinguishing the measurement outcomes can be indicated by a set of
values, called eigenvalues. As the variable $X$ or $Y$ takes a value of
$\omega^n \in S$, let the set of eigenvalues be the set $S$ such that
the operator is represented by $\hat{X} = \sum_{n=0}^{d-1} \omega^n
|n\rangle_{x x} \langle n|$.  Similarly, $\hat{Y} = \sum_{n=0}^{d-1}
\omega^n |n\rangle_{y y} \langle n|$. In this representation the
``observable'' operator $\hat{X}$ or $\hat{Y}$ is unitary \cite{Cerf02}.
Each measurement described is nondegenerate with all distinct
eigenvalues, hence called a maximal test \cite{Peres98}. We note that
such a unitary representation induces mathematical simplifications
without altering any physical results.

Consider the observable operator $\hat{X}$ that we obtain by applying
quantum Fourier transformation $\hat{Q}$ on the reference observable
$\hat{Z}=\sum_n \omega^n |n\rangle \langle n|$. For each eigenvalue
$\omega^n$, the eigenvector of $\hat{X}$ is thus given by
\begin{eqnarray}
  \label{eq:loofx}
  |n\rangle_x = \hat{Q} |n\rangle = \frac{1}{\sqrt{d}}\sum_{m=0}^{d-1}
   \omega^{nm}|m\rangle.
\end{eqnarray}
The observable $\hat{X}$ can be represented in terms of the reference
basis set $\{|n\rangle\}$ by
\begin{eqnarray}
  \label{eq:obox}
  \hat{X} = \sum_{n=0}^{d-1} |n\rangle \langle n+1|,
\end{eqnarray}
where we used the convention that $|n\rangle \equiv |n \mod d \rangle$
and thus $|d\rangle \equiv |0\rangle$. The operator $\hat{X}$ performs a
periodic shift operation on a basis vector:
\begin{eqnarray}
  \label{eq:obxobv}
  |n+1\rangle \rightarrow |n\rangle~~\mbox{and}~~|0\rangle \rightarrow
   |d-1\rangle. 
\end{eqnarray}
Then, the generalized GHZ state $|\psi\rangle$ in Eq.~(\ref{eq:ghzs}) is
the eigenstate of the composite observable $\hat{v}_0=\hat{X}\otimes
\hat{X} \otimes \hat{X}$ with the unit eigenvalue as
\begin{eqnarray}
  \label{eq:evefxxx}
  \hat{X} \otimes \hat{X}\otimes \hat{X}|\psi\rangle 
  &=& \frac{1}{\sqrt{d}}\sum_{n=0}^{d-1} |n-1, n-1,
  n-1\rangle \nonumber \\
  &=& |\psi\rangle. 
\end{eqnarray}
We note that $\hat{X}$ has the form of Eq.~(\ref{eq:lust}) as
$\hat{v}_0$ is also a symmetry operator for which $|\psi\rangle$ remains
invariant.

By using the symmetric operations (\ref{eq:lut}) for the generalized GHZ
state $|\psi\rangle$, we can construct other concurrent observables from
the composite observable $\hat{v}_0$.  Such a typical unitary operator
$\hat{V}_1 = \hat{U}_1 \otimes \hat{U}_2 \otimes \hat{U}_2$ where
$\hat{U}_\alpha$ are given with $g(n) = n$, $f_1(n)=(d-1)n$, and
$f_2(n)=n/2$ in Eq.~(\ref{eq:lut}).  Note that $\hat{V}_1$ satisfies the
condition~(\ref{eq:pfluoiqs}) as $f_1(n) + 2f_2(n) = dn \equiv 0
\mod d$ for all $n$ and it leaves the state $|\psi\rangle$ invariant.
The observable obtained is $\hat{v}_1=\hat{V}_1 \hat{v}_0 \hat{V}_1^\dag
= \omega \hat{X} \otimes \hat{Y} \otimes \hat{Y}$.  Here the observable
operator $\hat{Y}$ is of the measurement $Y$. For each eigenvalue
$\omega^n$, the eigenvector of $\hat{Y}$ is given by applying a phase
shift operation $\hat{U}_2$ on $|n\rangle_x$:
\begin{eqnarray}
  \label{eq:loofy}
  |n\rangle_y = \hat{U}_2 |n\rangle_x = \frac{1}{\sqrt{d}}
  \sum_{m=0}^{d-1} \omega^{(n+\frac{1}{2})m} |m\rangle.
\end{eqnarray}
The operator $\hat{Y}$ can be written similarly to Eq.~(\ref{eq:obox})
by
\begin{eqnarray}
  \label{eq:oboy}
  \hat{Y} = \omega^{-\frac{1}{2}} \left(\sum_{n=0}^{d-2} |n\rangle \langle n+1| -
  |d-1\rangle \langle 0|\right).
\end{eqnarray}
Contrary to $\hat{X}$, the operator $\hat{Y}$ performs an antiperiodic
shift operation with a phase shift $\omega^{-1/2}$: 
\begin{eqnarray}
  \label{eq:obxobv}
  |n+1\rangle \rightarrow
  \omega^{-\frac{1}{2}}|n\rangle~~\mbox{and}~~|0\rangle \rightarrow 
   -\omega^{-\frac{1}{2}}|d-1\rangle. 
\end{eqnarray}
In the similar manner, we obtain the other two concurrent observables,
$\hat{v}_2=\omega \hat{Y} \otimes \hat{X} \otimes \hat{Y}$ and
$\hat{v}_3=\omega \hat{Y} \otimes \hat{Y} \otimes \hat{X}$ by applying
$\hat{V}_2 = \hat{U}_2 \otimes \hat{U}_1 \otimes \hat{U}_2$ and
$\hat{V}_3 = \hat{U}_2 \otimes \hat{U}_2 \otimes \hat{U}_1$,
respectively. The obtained three observables $\hat{v}_i$ respectively
satisfy
\begin{eqnarray}
  \label{eq:tclo}
  \hat{X} \otimes \hat{Y} \otimes \hat{Y} |\psi\rangle &=& \omega^{-1}
  |\psi\rangle, \nonumber \\ 
  \hat{Y} \otimes \hat{X} \otimes \hat{Y} |\psi\rangle &=& \omega^{-1}
  |\psi\rangle, \nonumber \\   
  \hat{Y} \otimes \hat{Y} \otimes \hat{X} |\psi\rangle &=& \omega^{-1}
  |\psi\rangle.
\end{eqnarray}
We note that the four concurrent observables $\hat{v}_i$ have a common
eigenstate of $|\psi\rangle$, even though they are mutually
incompatible, i.e., $[\hat{v}_i,\hat{v}_j] \ne 0$ for $i \ne j$.

Quantum mechanics allows that the concurrent observables $\hat{v}_i$ can
simultaneously be specified as far as the composite system is prepared
in the generalized GHZ state $|\psi\rangle$, as they satisfy
Eqs.~(\ref{eq:evefxxx}) and (\ref{eq:tclo}). In other words, all the
composite measurements $\hat{v}_i$ collapse the state $|\psi\rangle$ to
itself and the order of the measurements does not affect the result.
Nevertheless, the value of the local variable $X$ or $Y$ for each qudit
is revealed only by actually performing the measurement.

On the other hand, the local realistic description implies that the
local variables $X$ and $Y$ are elements of physical reality and the
values of the local variables $X$ and $Y$ are predetermined before the
measurements, contrary to the quantum mechanical description. We then
attempt to assign values to the variables $X_\alpha$ and $Y_\alpha$ for
each qudit $\alpha$.  This attempt converts Eqs.~(\ref{eq:tclo}) to the
algebraic equations that the predetermined variables $X_\alpha$ and
$Y_\alpha$ must obey:
\begin{eqnarray}
  \label{eq:abeqpva1}
  X_A Y_B Y_C &=& \omega^{-1}, \nonumber \\  
  Y_A X_B Y_C &=& \omega^{-1}, \nonumber \\  
  Y_A Y_B X_C &=& \omega^{-1}.
\end{eqnarray}
By definition $X_\alpha = \omega^{x_\alpha}$ and $Y_\alpha =
\omega^{y_\alpha}$, where $x_\alpha$ and $y_\alpha$ are integers, and
the above equations can be rewritten in a simpler form of
\begin{eqnarray}
  \label{eq:abeqpva}
  x_A + y_B + y_C &\equiv& -1 \mod d, \nonumber \\  
  y_A + x_B + y_C &\equiv& -1 \mod d, \nonumber \\  
  y_A + y_B + x_C &\equiv& -1 \mod d.
\end{eqnarray}
Summing these equations results in the ``local realistic condition'':
\begin{eqnarray}
  \label{eq:lrrfepr}
  x_A+x_B+x_C \equiv  -2(y_A+y_B+y_C)-3 \mod d.
\end{eqnarray}
For an even integer $d$, the right hand side of Eq.~(\ref{eq:lrrfepr})
is always an odd integer modulo $d$ for arbitrary $y_\alpha$. In other
words, for even $d$, there exist no integer solutions of $y=y_A+y_B+y_C$
to the equation $2y+3\equiv 0 \mod d$.  This is in contradiction to the
quantum expectation, from Eq.~(\ref{eq:evefxxx}),
\begin{eqnarray}
  \label{eq:qrfepr}
  x_A+x_B+x_C \equiv 0 \mod d.
\end{eqnarray}
Thus we prove the Bell theorem without statistical inequalities for an
arbitrary even dimensional tripartite system.  For $d=2$, in particular,
the observables $\hat{X}$ and $\hat{Y}$ respectively reduce
$\hat{\sigma}_x$ and $\hat{\sigma}_y$ with $\omega=-1$ and the
nonlocality is equivalent to that originally proposed by GHZ
\cite{GHZ89}.

\subsection{Genuine multi-dimensionality}
\label{sec:gmghz}

One may try to extend the GHZ nonlocality in two dimension \cite{GHZ89}
to higher dimensions by employing anticommuting observables
\cite{Cabello01}.  However, such an extension is a {\em de facto}
two-dimensional nonlocality \cite{Cerf02}. It is because two
anticommuting observables can always be represented by a direct sum of
two dimensional observables.

To confirm that the generalized GHZ nonlocality is genuinely
$d$-dimensional, we prove that it is impossible to represent the
observables $\hat{X}$ and $\hat{Y}$ by a direct sum of any
subdimensional observables. Suppose that the observable operator
$\hat{X}$ were block-diagonalizable by some similarity transformation
$\hat{S}$ such that $\hat{S}^{-1} \hat{X} \hat{S} = \hat{X}_1 \oplus
\cdots \oplus \hat{X}_N$. Suppose further that $\hat{S}$ could
simultaneously block-diagonalize the observable operator $\hat{Y}$:
$\hat{S}^{-1} \hat{Y} \hat{S} = \hat{Y}_1 \oplus \cdots \oplus
\hat{Y}_N$. Here $\hat{X}_i$ and $\hat{Y}_i$ are observables of $d_i$
dimension with $\sum_i d_i = d$. Then, it should hold $\mbox{Tr}
\hat{X}_i \hat{Y}_j = 0$ for $i\ne j$. In other words, there should
exist some eigenvectors $|n\rangle_x$ of $\hat{X}$ and $|m\rangle_y$ of
$\hat{Y}$ such that $0={}_x\langle
n|\hat{S}\hat{S}^{-1}|m\rangle_y={}_x\langle n|m\rangle_y$. However, no
such eigenvectors can exist because for every $n$ and $m$
\begin{eqnarray}
  \label{eq:pmdghz}
  \left| {}_x\langle n | m \rangle_y \right|^2 = 
  \frac{1}{d^2\sin^2[\frac{\pi}{d}(m-n+\frac{1}{2})]} > 0.
\end{eqnarray}
Therefore the generalized GHZ nonlocality is genuinely $d$-dimensional.

\subsection{Extension to multipartite systems}
\label{sec:etms}

The tripartite GHZ nonlocality can be extended to a general $N$-partite
and $d$ dimensional system where $N$ is an odd integer and $d$ an even
integer. This extension requires a set of $(N+1)$ concurrent
observables, which includes $\hat{X}^{\otimes N}$, $\hat{X}\otimes
\hat{Y}^{\otimes (N-1)}$, and its permutations, i.e.,
\begin{eqnarray}
  \label{eq:csocos}
  \hat{v}_0&=&\hat{X}\otimes \hat{X} \otimes \hat{X} \otimes \cdots \otimes \hat{X}
  \nonumber \\
  \hat{v}_1&=&\hat{X}\otimes \hat{Y} \otimes \hat{Y} \otimes \cdots \otimes \hat{Y}
  \nonumber \\
  \hat{v}_2&=&\hat{Y}\otimes \hat{X} \otimes \hat{Y} \otimes \cdots \otimes \hat{Y}
  \nonumber \\
  \hat{v}_3&=&\hat{Y}\otimes \hat{Y} \otimes \hat{X} \otimes \cdots \otimes \hat{Y}
  \nonumber \\
  &\vdots \nonumber \\
  \hat{v}_N&=&\hat{Y}\otimes \hat{Y} \otimes \hat{Y} \otimes \cdots \otimes \hat{X}.
\end{eqnarray}
Here $\hat{X}$ is given in Eq.~(\ref{eq:obox}), while $\hat{Y}$ is
modified by generalizing the local unitary operator $\hat{U}_2$ with
$f_2(n) = n/(N-1)$ from $f_2(n) = n/2$:
\begin{eqnarray}
  \label{eq:oboyN}
  \hat{Y} = \omega^{-\frac{1}{N-1}} \left(\sum_{n=0}^{d-2} |n\rangle \langle
  n+1| + \omega^{\frac{d}{N-1}} |d-1\rangle \langle 0|\right).
\end{eqnarray} 
The $N$-partite generalized GHZ state,
\begin{eqnarray}
  \label{eq:gghzons}
  |\psi\rangle = \frac{1}{\sqrt{d}} \sum_{n=0}^{d-1} |n,n,\dots,n\rangle,
\end{eqnarray}
is a common eigenstate of all the concurrent observables with the
eigenvalues 1 for $\hat{v}_0$ and $\omega^{-1}$ for the others
$\hat{v}_i$.  Following the argument as for the tripartite GHZ
nonlocality, we obtain the local realistic condition,
\begin{eqnarray}
  \label{eq:lrcfms}
  \sum_{\alpha=1}^N x_\alpha \equiv -\left(N-1\right)\sum_{\alpha=1}^N
  y_\alpha - N \mod d, 
\end{eqnarray}
where $x_\alpha$ and $y_\alpha$ come from the variables $X_\alpha =
\omega^{x_\alpha}$ and $Y_\alpha = \omega^{y_\alpha}$ for each qudit
$\alpha$. For {\em each} pair of odd $N$ and even $d$, this is in
contradiction to the quantum expectation, resulting from $\hat{v}_0$,
\begin{eqnarray}
  \label{eq:qcfms}
  \sum_{\alpha=1}^N x_\alpha \equiv 0 \mod d.
\end{eqnarray}
The pairs $(N,d)$ of odd $N$ and even $d$ include a particular element
of $(d+1,d)$. Our extension thus covers the previous works of
$d$-dimensional $(d+1)$-partite nonlocality \cite{Zukowski99,Cerf02}.

In order to test the generalized GHZ nonlocality, one may consider an
optical experiment of using multiport beam splitters and phase shifters,
similar to that by \.Zukowski and Kaszlikowski \cite{Zukowski99}.  It
was shown that all unitary operators on a qudit can be implemented by a
series of those linear optical devices \cite{Reck94}. Thus, one can
implement the local measurement bases for $\hat{X}$ and $\hat{Y}$ by
simply placing such optical devices before detectors.

\section{Remarks}
\label{sec:rem}

Our formulation of the generalized GHZ nonlocality is different from the
conventional approaches. First, it employs the concurrent observables
instead of compatible observables. Second, it releases the condition of
mutual complementarity between the local observables $\hat{X}$ and
$\hat{Y}$ \cite{Englert92,Lee03}. If the local observables $\hat{X}$ and
$\hat{Y}$ were mutually complementary, their eigenvectors would satisfy,
\begin{eqnarray}
  \label{eq:pmdghz2}
  \left| {}_x\langle n | m \rangle_y \right|^2 = \frac{1}{d}.
\end{eqnarray}
This is not the case as shown in Eq.~(\ref{eq:pmdghz}). These
differences enable a {\em tripartite} system to suffice for the
higher-than-two dimensional GHZ nonlocality, contrary to the previous
works that demand a $(d+1)$-partite system \cite{Zukowski99,Cerf02}.
This work will encourage to study the nonlocality for more general
systems.

In summary, we presented the genuinely multi-dimensional and
multipartite GHZ nonlocality.  The proof of nonlocality was based on the
concurrent observables that are incompatible but still have a common
eigenstate of the generalized GHZ state.

\acknowledgments

The authors thank M. \.Zukowski for bringing our attention to
Ref.~\cite{Kaszlikowski02} and W. Son and Y. W. Kim for helpful
discussions.  This work was supported by the Korean Ministry of Science
and Technology through Quantum Photonic Science Research Center. MSK
thanks to the UK Engineering and Physical Science Research Council for
financial support and KRF (2003-070-C00024).

\end{document}